\documentclass[aps,prb,twocolumn]{revtex4-1}

\usepackage[pdftex]{graphics}
\usepackage{amssymb,amsmath}
\usepackage{nicefrac}
\usepackage{color}

\begin{document}
\def\bra#1{\left<{#1}\right|}
\def\ket#1{\left|{#1}\right>}
\def\expval#1#2{\bra{#2} {#1} \ket{#2}}

\title{An efficient quantum mechanical method for radical pair recombination reactions}
\author{Alan M.~Lewis}
\affiliation{Department of Chemistry, University of Oxford, Physical and Theoretical Chemistry Laboratory, South Parks Road, Oxford, OX1 3QZ, UK}
\author{Thomas P.~Fay}
\affiliation{Department of Chemistry, University of Oxford, Physical and Theoretical Chemistry Laboratory, South Parks Road, Oxford, OX1 3QZ, UK}
\author{David E. Manolopoulos}
\affiliation{Department of Chemistry, University of Oxford, Physical and Theoretical Chemistry Laboratory, South Parks Road, Oxford, OX1 3QZ, UK}

\begin{abstract}
The standard quantum mechanical expressions for the singlet and triplet survival probabilities and product yields of a radical pair recombination reaction involve a trace over the states in a combined electronic and nuclear spin Hilbert space. If this trace is evaluated deterministically, by performing a separate time-dependent wavepacket calculation for each initial state in the Hilbert space, the computational effort scales as $\mathcal{O}(Z^2\log Z)$, where $Z$ is the total number of nuclear spin states. Here we show that the trace can also be evaluated stochastically, by exploiting the properties of spin coherent states. This results in a computational effort of $\mathcal{O}(MZ\log Z)$, where $M$ is the number of Monte Carlo samples needed for convergence. Example calculations on a strongly-coupled radical pair with $Z>10^6$ show that the singlet yield can be converged to graphical accuracy using just $M=200$ samples, resulting in a speed up by a factor of $>5000$ over a standard deterministic calculation. We expect that this factor will greatly facilitate future quantum mechanical simulations of a wide variety of radical pairs of interest in chemistry and biology.
\end{abstract}

\maketitle

\section{Introduction}

Spin dynamics play an important role in many systems of biological and chemical interest, such as magnetoreceptors,\cite{Maeda08,Rodgers09b,Mouritsen12} organic semiconductors,\cite{Lupton10, Nguyen10, Ehrenfreund12} and molecular wires.\cite{Weiss04, Tauber06, Wasielewski06} The spin-selective behaviour exhibited by these systems is often explained using the radical pair model, in which two coupled electron spins interact with a number of hyperfine-coupled nuclear spins. When this number exceeds twenty or so,  standard quantum mechanical calculations become cripplingly expensive, and while in some cases semiclassical approximations have been shown to be accurate,\cite{Schulten78,Manolopoulos13,Lewis14,Lawrence16} this cannot always be relied on. As a result, there is a clear need for a more efficient quantum mechanical method for describing the spin dynamics of radical pairs.

\begin{figure}[t]
\centering
\resizebox{0.55\columnwidth}{!} {\includegraphics{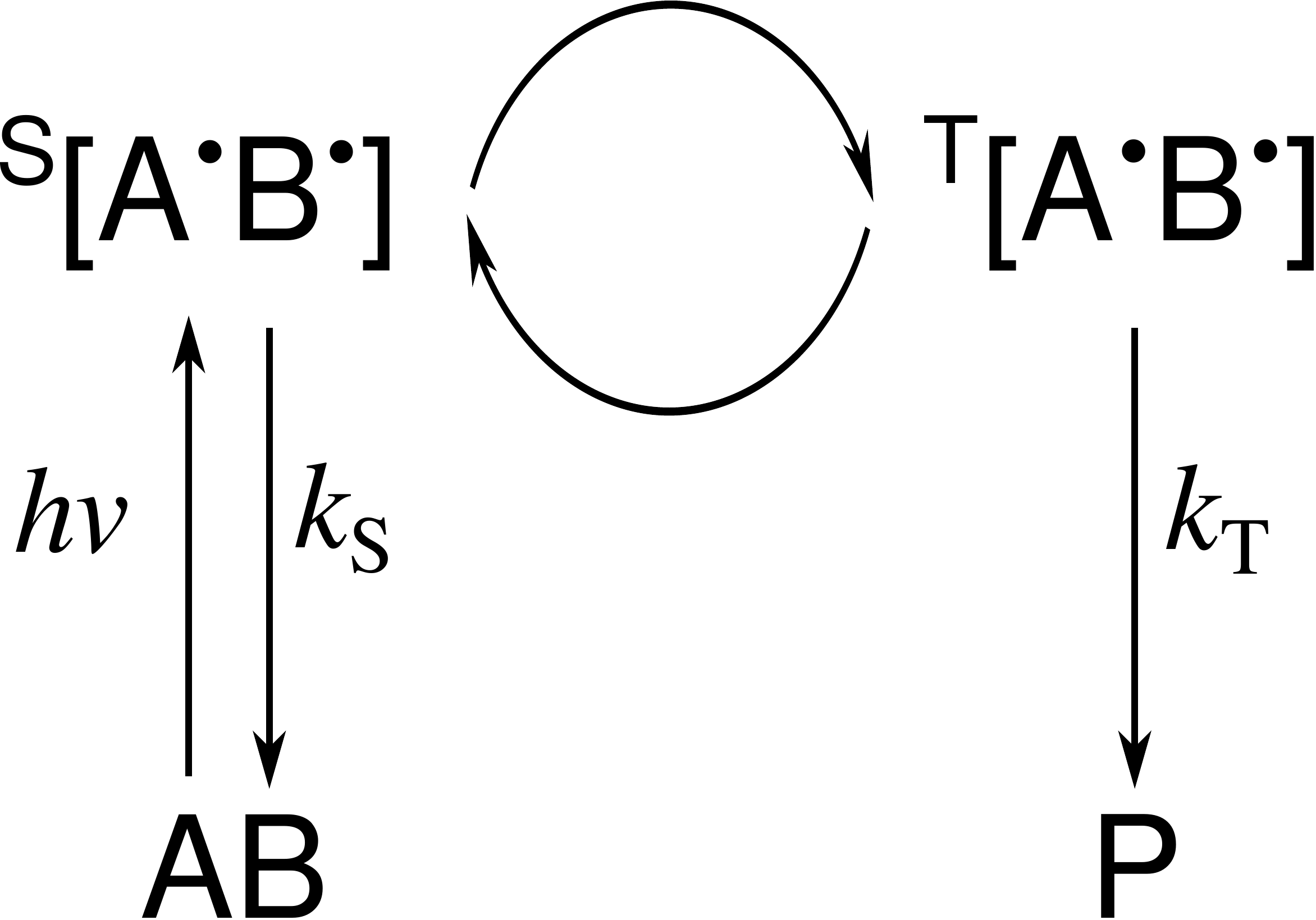}}
\caption{An idealised radical pair recombination reaction. The superscripts S and T label the singlet and triplet states of the radical pair respectively; $k_{\rm S}$ and $k_{\rm T}$ are the recombination rate constants for these states. The curved arrows represent hyperfine-mediated intersystem crossing between the singlet and triplet states.}
\label{RPR}
\end{figure}

A simplified radical pair reaction scheme is given in Figure \ref{RPR}. Photoexcitation of an organic precursor molecule produces a radical pair in the singlet state, which can then undergo intersystem crossing to form the triplet state. The singlet and triplet radical pairs can each recombine, in general at different rates and forming different products. The intersystem crossing is mediated by hyperfine interactions between the electron and nuclear spins in each radical, and as a result is affected by the application of an external magnetic field. This in turn can modify the lifetime of the radical pair, and the singlet and triplet recombination yields. The resulting magnetic field effects (MFEs)\cite{Rodgers09,Volk95,Timmel98} are often measured in order to probe the spin dynamics of the radical pair, but it is usually not possible to extract the spin-specific rate constants $k_{\rm S}$ and $k_{\rm T}$ from the experiments alone. More insight can often be gained by comparing the experimental results to simulations.

The quantum dynamics of the radical pair model is well understood, and simulations of small spin systems can be done routinely.\cite{Steiner89} However, simulating realistic systems is often challenging, due to the exponential scaling of the size of the Hilbert space with the number of nuclear spins. Previous theoretical calculations have been performed either by neglecting smaller hyperfine couplings to reduce the size of the Hilbert space,\cite{Cintolesi03,Efimova08} or by using semiclassical methods to avoid the exponential scaling.\cite{Schulten78,Manolopoulos13,Lewis14,Lawrence16} However, neither approach is entirely reliable for systems with $N \sim 20$ nuclear spins, which are commonplace in chemical and biological systems. It is not always safe to assume that weakly-coupled nuclear spins make no contribution to the spin dynamics, and semiclassical approximations are incapable of capturing subtle quantum mechanical effects such as the recently-discovered \lq\lq quantum needle" of the avian magnetic compass.\cite{Hiscock16}

We shall therefore now present a new method that dramatically improves the efficiency of quantum mechanical spin dynamics simulations, allowing the treatment of significantly larger radical pairs. To illustrate this, we shall calculate the singlet yield of a strongly-coupled radical pair involving 20 nuclear spins, over a wide range of magnetic field strengths. This calculation would be quite impractical using a standard quantum mechanical method, and while various semiclassical approximations are applicable to the problem we shall demonstrate that they are only qualitatively accurate.

\section{Standard Theory}

The general form of the Hamiltonian that governs the spin dynamics of a radical pair tumbling in solution includes isotropic Zeeman and hyperfine interactions and an exchange coupling between the two electron spins:
\begin{equation}
\begin{gathered}
\hat{H} = \hat{H}_1 + \hat{H}_2 + 2J\,\hat{\bf S}_1\cdot\hat{\bf S}_2, \\
\hat{H}_i = \boldsymbol{\omega}\cdot\hat{\bf S}_i+\sum_{k=1}^{N_i} a_{ik}\,\hat{\bf I}_{ik}\cdot\hat{\bf S}_i.
\label{Hamiltonian}
\end{gathered}
\end{equation}
Here $\boldsymbol{\omega} = -\gamma{\bf B}$, where $\gamma$ is the gyromagnetic ratio of the electron and ${\bf B}$ is the applied magnetic field. $a_{ik}$ is the hyperfine coupling constant between the $k$th nuclear spin and the electron spin in radical $i$; $\hat{\bf I}_{ik}$ and $\hat{\bf S}_i$ are the corresponding nuclear and electron spin operators, $N_i$ is the number of nuclear spins in the radical, and $J$ is the exchange coupling constant. Note that we have neglected the comparatively weak nuclear spin Zeeman interactions, { we have assumed for simplicity that the electrons in the two radicals have the same gyromagnetic ratio}, and we are working in a unit system in which $\hbar=1$. 

The recombination of the radical pair is conventionally modelled using the Haberkorn operator,\cite{Haberkorn76,Ivanov10}
\begin{equation}
\hat{K} = {k_{\rm S}\over 2}\hat{P}_{\rm S}+{k_{\rm T}\over 2}\hat{P}_{\rm T},
\end{equation}
where
\begin{equation}
\hat{P}_{\rm S} = {1\over 4}\hat{\bf 1}-\hat{\bf S}_1\cdot\hat{\bf S}_2
\end{equation}
\begin{equation}
\hat{P}_{\rm T} = {3\over 4}\hat{\bf 1}+\hat{\bf S}_1\cdot\hat{\bf S}_2
\end{equation}
are the projection operators onto the singlet and triplet electronic subspaces, and $k_{\rm S}$ and $k_{\rm T}$ are the first order rate constants for recombination of singlet and triplet states respectively. The evolution of the density operator $\hat{\rho}(t)$ which describes the radical pair is then governed by the quantum Liouville equation\cite{Haberkorn76,Neumann55}
\begin{equation}
{d\over dt}\hat{\rho}(t) = -i[\hat{H},\hat{\rho}(t)]-\{\hat{K},\hat{\rho}(t)\},
\label{Liouville}
\end{equation}
where $[\hat{A},\hat{B}]$ is the commutator and $\{\hat{A},\hat{B}\}$ the anticommutator of $\hat{A}$ and $\hat{B}$. This differential equation is satisfied by
\begin{equation}
\hat{\rho}(t) = e^{-i\hat{H}t-\hat{K}t}\,\hat{\rho}(0)\,e^{+i\hat{H}t-\hat{K}t}.
\end{equation}

The ensemble average of an observable corresponding to an operator $\hat{A}$ at time $t$ is given by
\begin{equation}
A(t) = {\rm tr}\left[\hat{\rho}(t)\hat{A}(0)\right]
\end{equation}
or, using the invariance of a trace to cyclic permutation,
\begin{equation}
A(t) = {\rm tr}\left[\hat{\rho}(0)\hat{A}(t)\right],
\label{trace}
\end{equation}
where
\begin{equation}
\hat{A}(t) = e^{+i\hat{H}t-\hat{K}t}\,\hat{A}\,e^{-i\hat{H}t-\hat{K}t}.
\end{equation}
We may evaluate the trace in Eq.~\eqref{trace} in the basis
\begin{equation}
\{\mathcal{B}\} = \{\ket{\Theta} \otimes \ket{{\bf M}_1} \otimes \ket{{\bf M}_2}\}.
\end{equation}
Here $\ket{\Theta}$ is the electronic spin state, in which $\Theta$ can take the values ${\rm S}$, ${\rm T_+}$, ${\rm T_0}$ and ${\rm T_-}$, representing the singlet and triplet states respectively. $\ket{{\bf M}_i}$ is the nuclear spin state of radical $i$, given by
\begin{equation}
\ket{{\bf M}_i} = \ket{M_{i1}} \otimes \ket{M_{i2}} \otimes \dots \otimes \ket{M_{iN_i}},
\end{equation}
where $M_{ik}$ is the projection of the $k$th nuclear spin in the radical onto the $z$ axis. In this basis, Eq.~\eqref{trace} becomes
\begin{equation}
A(t) = \sum_\Theta\sum_{{\bf M}_1}\sum_{{\bf M}_2}\expval {\hat{\rho}(0)\hat{A}(t)}{\Theta,{\bf M}_1,{\bf M}_2}
\label{traceinbasis}
\end{equation}
where $\sum_{{\bf M}_i}$ indicates the sum over all possible nuclear spin states $\ket{{\bf M}_i}$.

In the case of photoexcited radical pairs formed in the singlet state, $\hat{\rho}(0) = \hat{P}_{\rm S} / Z$, where $Z = \Pi_{i,k} (2I_{ik} + 1)$ is the total number of nuclear spin states in the radical pair. The sum over $\Theta$ in Eq.~\eqref{traceinbasis} can then be evaluated immediately: the only term that survives is that with $\Theta = {\rm S}$. Defining
\begin{equation}
\ket{{\rm S},{\bf M}_1,{\bf M}_2;t} = e^{-i\hat{H}t-\hat{K}t} \ket{{\rm S},{\bf M}_1,{\bf M}_2},
\label{prop}
\end{equation}
the trace becomes
\begin{equation}
A(t) = \frac{1}{Z}\sum_{{\bf M}_1}\sum_{{\bf M}_2}\expval {\hat{A}}{{\rm S},{\bf M}_1,{\bf M}_2;t}.
\label{expandedtrace}
\end{equation}
Therefore, the ensemble average $A(t)$ can be evaluated by propagating $Z$ wavepackets with orthogonal initial nuclear spin states and calculating the expectation value of $\hat{A}$ in each state at time $t$.

The wavepacket propagation in Eq.~\eqref{prop} can be performed using the short iterative Arnoldi (SIA) method.\cite{Pollard94} This is a generalisation of the short iterative Lanczos (SIL) method\cite{Park86} which allows for propagators generated from non-Hermitian operators, as is the case here. The sparsity of the Hamiltonian means that Hamiltonian-wavepacket multiplications require only $\mathcal{O} (Z\log{Z})$ operations,\cite{count} and combined with the adaptive time step of the SIA method, this leads to very efficient wavepacket propagation. As a result, the computational time required to evaluate Eq.~\eqref{expandedtrace} scales as $\mathcal{O}(Z^2\log{Z})$. In order to improve this scaling, either the efficiency of the wavepacket propagation must be improved, or fewer propagations must be carried out. We will now demonstrate that the latter is easily achieved if the trace in Eq.~\eqref{trace} is evaluated using Monte Carlo integration over the coherent spin state basis of the nuclear spins.

\section{Spin coherent states}

For a spin with quantum number $J$, a coherent spin state $\ket{\Omega_J}$ is obtained by rotating the quantisation axis of the $\ket{J,J}$ eigenstate of $\hat{J}_z$ to lie in the direction $\Omega = (\theta,\phi)$. It is defined as\cite{Radcliffe71}
\begin{equation}
\ket{\Omega_J} = (\cos{\theta/2})^{2J}\,\exp\{\tan{(\theta/2)}\,e^{i\phi}\hat{J}_-\}\ket{J,J},
\end{equation}
where $0 \leqslant \theta \leqslant \pi$ and $0 \leqslant \phi < 2\pi$. The expectation value of the angular momentum operator ${\hat{\bf J}}$ in state $\ket{\Omega_J}$ corresponds to a classical vector of length $J$ in the direction $\Omega$:
\begin{equation}
\begin{gathered}
\expval{\hat{\bf J}}{\Omega_J} = J{\bf n}(\Omega), \\
{\bf n}(\Omega) = (\sin{\theta}\cos{\phi},\sin{\theta}\sin{\phi},\cos{\theta})^\top.
\end{gathered}
\end{equation}

The states $\{\ket{\Omega_J}\}$ form an overcomplete set, with the completeness relation\cite{Radcliffe71}
\begin{equation}
\begin{aligned}
\hat{1}& = \frac{2J+1}{4\pi}\int {\rm d}\Omega \ket{\Omega_J} \bra{\Omega_J}.
\end{aligned}
\label{unit}
\end{equation}
Inserting equation Eq.~\eqref{unit} into Eq.~\eqref{expandedtrace}, we find that the sum over ${\bf M}_1$ and ${\bf M}_2$ can be replaced by a $2N$ dimensional integral over the coherent spin states of each nuclear spin,
\begin{equation}
A(t) = \frac{1}{(4\pi)^N}\int {\rm d}{\bf \Omega}_1 \int {\rm d}{\bf \Omega}_2 \expval {\hat{A}}{{\rm S},{\bf \Omega}_1,{\bf \Omega}_2;t},
\label{MC}
\end{equation}
where $N=N_1+N_2$ and
\begin{equation}
\ket{{\bf \Omega}_i} = \ket{\Omega_{i1}} \otimes \ket{\Omega_{i2}} \otimes \dots \otimes \ket{\Omega_{iN_i}}.
\end{equation}
Here $\ket{\Omega_{ik}}$ is a coherent spin state of the $k$th nuclear spin of radical $i$, and $\int$d${\bf \Omega}_i$ indicates the integral over all possible orientations of each of these coherent spin states. Note that the factors of $(2J+1)$ in Eq.~\eqref{unit} cancel with the factor of $1 / Z$ in Eq.~\eqref{expandedtrace}.

Eq.~\eqref{MC} provides an expression for the ensemble average of an observable in terms of a product of integrals over coherent spin states. These integrals, and hence $A(t)$, may be evaluated by Monte Carlo sampling the directions of the initial nuclear spin states $\ket{\Omega_{ik}}$ from the surfaces of their respective spheres. If the number of samples $M$ required to converge the integrals is significantly smaller than $Z$, the computational cost of calculating $A(t)$ will be greatly reduced compared with a deterministic evaluation of Eq.~(14).

\begin{table} [t]
\centering
\begin{tabular} { c  c  c }
$k$ & $a_{1k}/\gamma$ (mT) & $a_{2k}/\gamma$ (mT) \\ \\
\hline
\hline\\
1 & -0.999985 & -0.232996 \\
2 & -0.736925 & 0.0388327 \\
3 &  0.511210 &  0.661931 \\
4 &-0.0826998 & -0.930856 \\
5 & 0.0655341 & -0.893077 \\
6 & -0.562082 & 0.0594001 \\
7 & -0.905911 &  0.342299 \\
8 &  0.357729 & -0.984604 \\
9 &  0.358593 & -0.233169 \\
10 & 0.869386 & -0.866316 \\
\end{tabular}
\caption{Hyperfine coupling constants in a model radical pair.}
\label{Table}
\end{table}

\section{A Model Problem}

In order to investigate this, we have calculated the singlet yield of a particularly demanding radical pair in which the hyperfine, Zeeman, and electronic exchange interactions were all chosen to have similar orders of magnitude (the most strongly coupled scenario). This singlet yield is given by 
\begin{equation}
\Phi_{\rm S} = k_{\rm S} \int_0^{\infty} {\rm P_S}(t) \, {\rm d}t,
\end{equation}
where ${\rm P_S}(t)$ is the ensemble average of the singlet probability at time $t$, obtained by substituting $\hat{A} = \hat{P}_{\rm S}$ into Eq.~\eqref{MC}. In our model system, each electron was coupled to ten $I=1/2$ nuclear spins with randomly generated hyperfine coupling constants uniformly distributed between $-1 < a_{ik} < 1$ mT; these hyperfine constants are listed for completeness in Table \ref{Table}. The exchange coupling constant was taken to be $J  / \gamma = 1.75$ mT, comparable to the effective hyperfine field strength 
\begin{equation}
B_{{\rm hyp},i} = \sqrt{\sum_{k=1}^{N_i} a_{ik}^2I_{ik}(I_{ik}+1)}
\end{equation}
in each radical ($B_{{\rm hyp},1}=1.72$ mT and $B_{{\rm hyp},2}=1.74$ mT), and the recombination rate constants were taken to be $k_{\rm S}  / \gamma = 0.1$ mT and $k_{\rm T} / \gamma = 1.0$ mT.

\section{Results and Discussion}

\begin{figure}[t]
\centering
\resizebox{0.75\columnwidth}{!} {\includegraphics{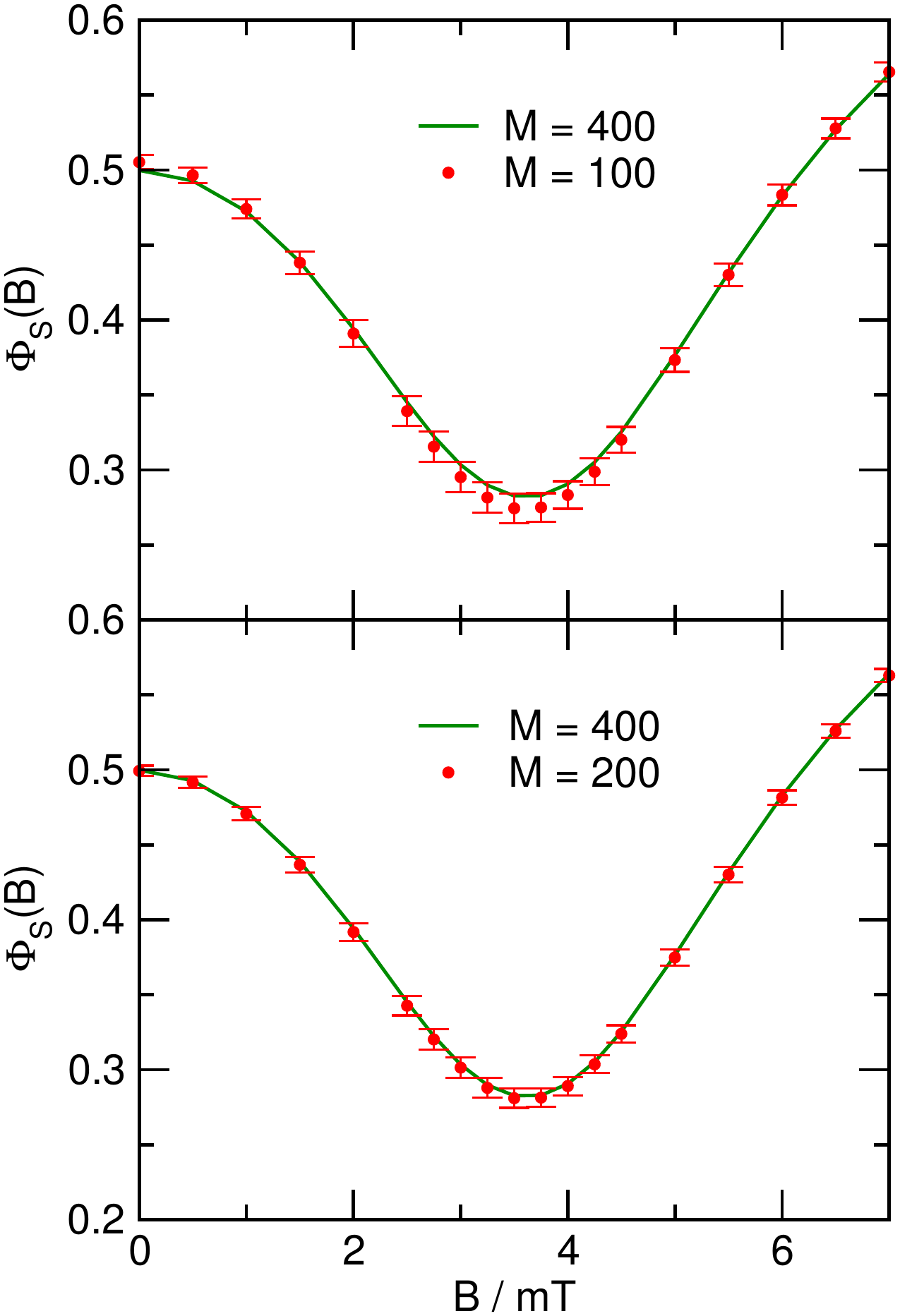}}
\caption{ The singlet yield of the model radical pair as a function of the applied magnetic field strength $B$, averaged over $M$ initial coherent nuclear spin states. The two panels compare the results of calculations with $M=100$ and $M=200$ with those of a well converged calculation with $M=400$. The statistical error bars in the $M=100$ and $M=200$ results are $\pm\epsilon$ in Eq.~(23), with $\sigma$ estimated from the $M$ samples of $\Phi_{\rm S}(B)$ at each magnetic field strength $B$.}
\label{Convergence}
\end{figure}

Figure \ref{Convergence} shows how the singlet singlet yield of this model radical pair converges with increasing $M$ over a wide range of magnetic field strengths. The results are converged to graphical accuracy at all field strengths with only $M = 200$ samples. Since the total number of nuclear spin states in the radical pair is $Z =$ 1,048,576, this stochastic calculation was $> 5000$ times faster than the equivalent deterministic calculation. 

\begin{figure}[b]
\centering
\resizebox{0.75\columnwidth}{!} {\includegraphics{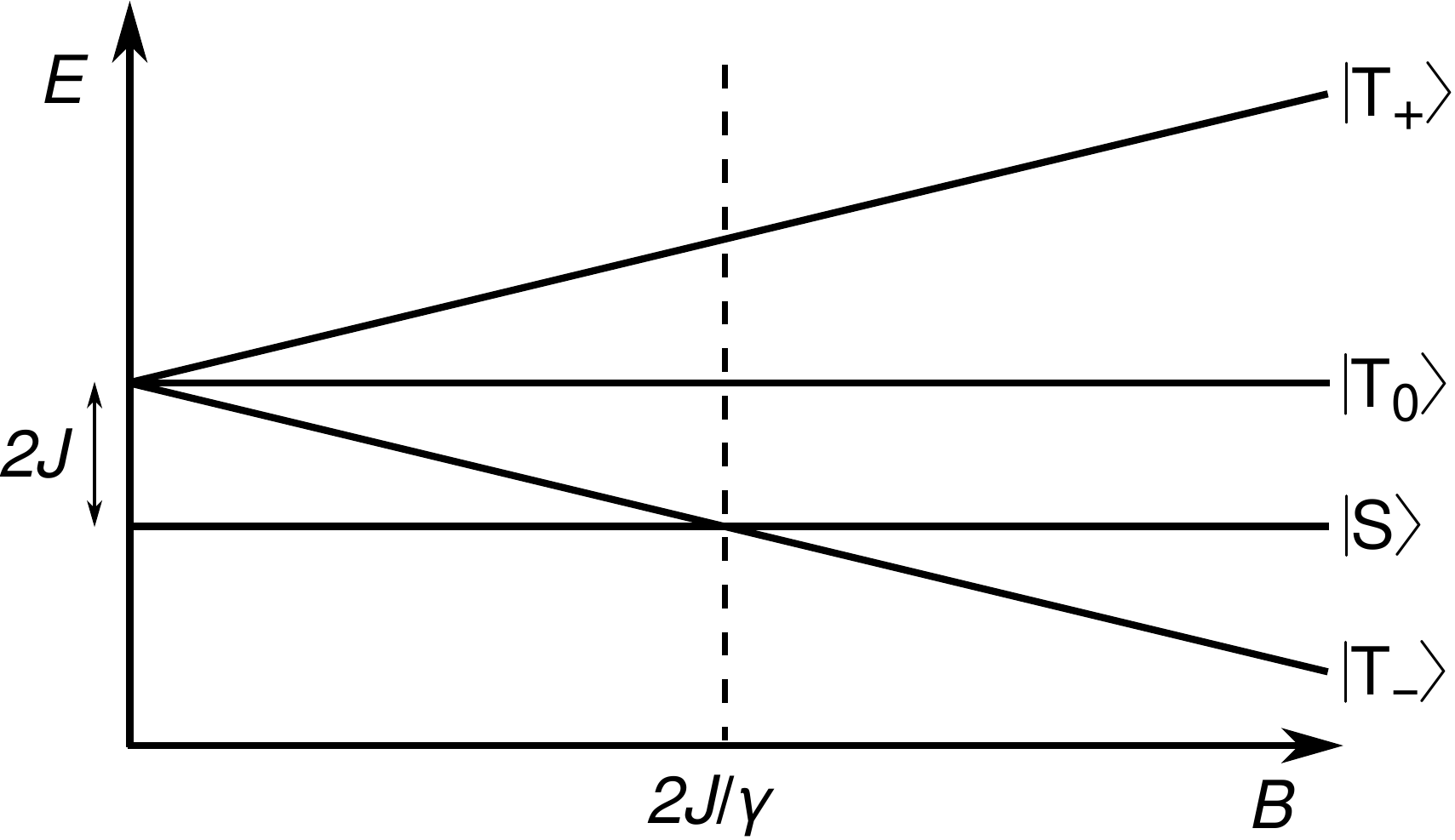}}
\caption{The energy levels of the spin states of the radical pair as a function of the strength of the applied magnetic field, $B$. The effect of hyperfine interactions has been neglected.}
\label{Energies}
\end{figure}

The form of the MFE on the singlet yield shown in Fig.~\ref{Convergence} can be understood by considering the energy levels of the electronic spin states of the radical pair, shown schematically in Figure \ref{Energies}. At both zero and high fields, the singlet state is energetically separated from the triplet states, which leads to slow intersystem crossing. But when $\omega = 2J$, the singlet state is degenerate with the ${\rm T}_{-}$ state, and hyperfine mediated intersystem crossing becomes much more efficient. Since the radical pair is formed in the singlet state, any intersystem crossing will reduce the singlet yield. Therefore, a minimum is observed in the the singlet yield when $\omega = 2J$. This also explains the observation { from the statistical error bars in the top panel of} Fig.~\ref{Convergence} that the convergence of the Monte Carlo integration is slowest around the minimum in the singlet yield. Away from resonance, the hyperfine interactions play only a small role in the spin dynamics, and the singlet yield obtained from each individual wavepacket does not vary significantly between Monte Carlo samples. However, when the intersystem crossing is fast, the singlet yield depends more strongly on the initial nuclear spin state, and more samples are required for convergence.

\begin{figure}[t]
\resizebox{0.7\columnwidth}{!} {\includegraphics{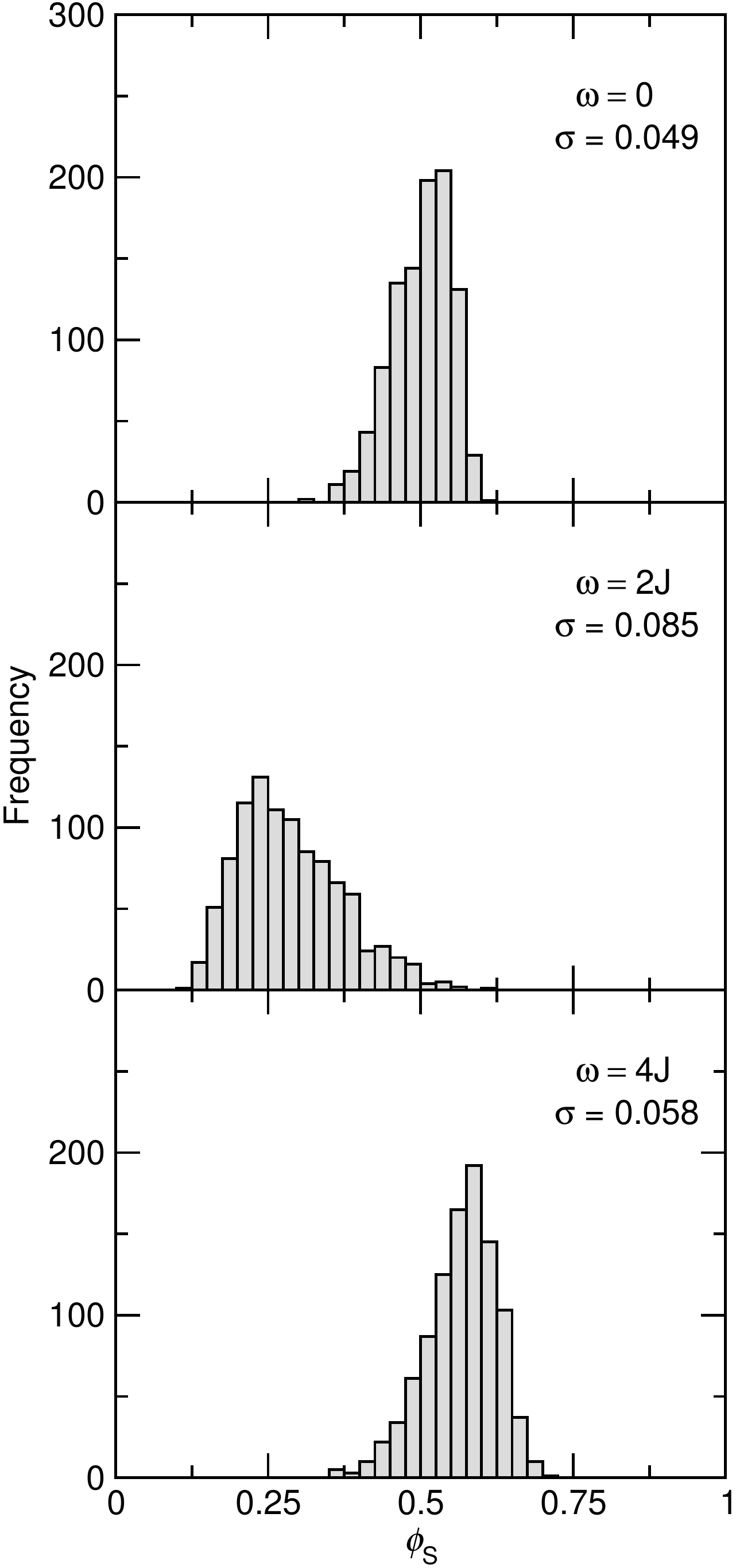}}
\caption{Histograms of 1000 single wavepacket singlet yields, $\phi_{\rm S}$, of the model radical pair at three different applied magnetic field strengths. The standard deviations of the distributions are also given.}
\label{Hist}
\end{figure}

The field dependent sensitivity of the spin dynamics to the initial nuclear spin state is illustrated by the distribution of singlet yields obtained from individual wavepackets,
\begin{equation}
\phi_{\rm S}({\bf \Omega}_1,{\bf \Omega}_2) = 
k_{\rm S}\int_0^{\infty} \expval {\hat{P}_{\rm S}}{{\rm S},{\bf \Omega}_1,{\bf \Omega}_2;t}{\rm d}t. 
\end{equation}
Figure \ref{Hist} shows a histogram of 1000 evaluations of this single wavepacket contribution to the singlet yield at three different magnetic field strengths. This highlights the increased variation in $\phi_{\rm S}$ when intersystem crossing becomes more efficient. Nevertheless, even when $\omega = 2J$, the number of coherent spin state samples required to converge the ensemble-averaged singlet yield is still far smaller than the total number of nuclear spin states, $Z$.

\begin{figure}[t]
\resizebox{0.75\columnwidth}{!} {\includegraphics{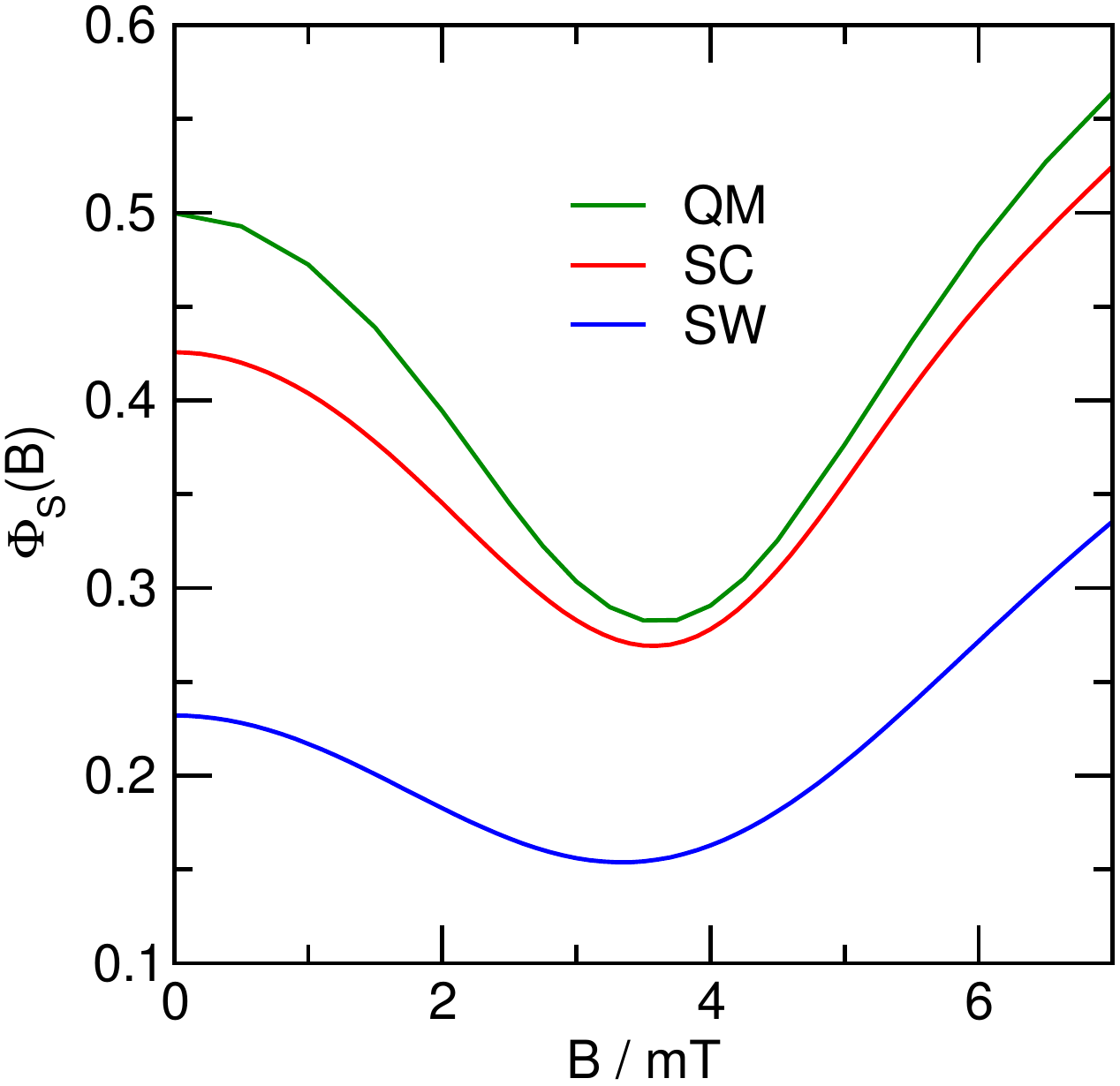}}
\caption{The singlet yield of the model radical pair as a function of magnetic field strength $B$, calculated using the present quantum mechanical (QM) approach, the semiclassical (SC) method presented in Refs.~\onlinecite{Manolopoulos13,Lewis14}, and a variant of Schulten-Wolynes\cite{Schulten78} (SW) theory outlined in Ref.~\onlinecite{Lawrence16}.}
\label{Methods}
\end{figure}

This analysis also explains why stochastic integration over coherent spin states is so efficient. Since the singlet yield is bounded between zero and one, the maximum possible standard deviation in the distribution of $\phi_{\rm S}$ is $\sigma = 1/2$. In practice, the distribution is confined to a smaller range, and so has a smaller standard deviation. When performing a Monte Carlo integration with random sampling, the standard deviation is related to the standard (statistical) error in the integral, $\epsilon$, and the number of samples, $M$, by\cite{Sobol98}
\begin{equation}
\epsilon = \frac{\sigma}{\sqrt{M}}.
\label{conv}
\end{equation}
Therefore, the necessarily small standard deviation in the distribution of singlet yields means that only a small number of samples is required for convergence. We expect this observation to be general, since the observables of interest in radical pair spin dynamics calculations are invariably probabilities or yields, which must be bounded between zero and one. As a result, the stochastic approach should be applicable to a wide range of radical pair observables. While the precise number of samples required to obtain converged results will doubtless vary from system to system, it seems likely that for radical pairs with $N > 10$ nuclear spins, $M$ will be significantly smaller than the number of nuclear spin states.

In order to establish whether a fully quantum mechanical simulation is necessary to describe systems of this type, we have also applied two semiclassical methods to this model problem. Figure \ref{Methods} compares the converged quantum mechanical (QM) results with those obtained using the original semiclassical theory of Schulten and Wolynes (SW),\cite{Schulten78} adapted to the calculation of radical pair singlet yields as described in Ref.~\onlinecite{Lawrence16}, and an improved semiclassical (SC) theory that takes the nuclear spin precession into account.\cite{Manolopoulos13,Lewis14} Unsurprisingly, the improved SC theory performs better than SW theory { across the whole range of magnetic field strengths}. However, neither method is quantitatively accurate. Clearly, in this case, where the hyperfine interactions, exchange coupling, and triplet recombination rate are all of a similar magnitude, semiclassical models are not accurate enough to capture the fine details of the spin dynamics.

Finally, as an alternative to the method presented here, we have considered the possibility of stochastically sampling ${\bf M}_1$ and ${\bf M}_2$ in order to evaluate the sums in Eq.~\eqref{expandedtrace}. We found that the convergence was much slower than when sampling coherent spin states. The reason for this lies in the fact that the Hamiltonian in Eq.~\eqref{Hamiltonian} commutes with the total spin projection operator, $\hat{J}_z = \hat{S}_{1z} + \hat{S}_{2z} + \sum_{i,k} \hat{I}_{ikz}$. This divides the Hilbert space into sectors of different $M_J$, which all contribute independently to $\Phi_{\rm S}$. Since the state $\ket{{\rm S},{\bf M}_1,{\bf M}_2}$ is an eigenstate of $\hat{J}_z$, each choice of ${\bf M}_1$ and ${\bf M}_2$ samples only a single sector of the total Hilbert space. By contrast, each coherent spin state includes some contribution from every $M_J$ sector in the space. If anisotropic hyperfine coupling or dipolar coupling were introduced, the Hamiltonian would no longer commute with $\hat{J}_z$, and so we expect that the convergence of results obtained by stochastically sampling the eigenstates of $\hat{J}_z$ would be faster. Nonetheless, we would still expect the results from sampling coherent spin states to converge at least as quickly.

\section{Conclusions and future work}

In this paper, we have demonstrated that it is possible to accurately simulate the spin dynamics of radical pairs containing all physically relevant couplings between the electron spins and at least 20 nuclear spins. This is the typical size of radical pair that arises in many interesting problems in chemistry and biology. We have also shown that semiclassical methods are not always quantitatively accurate for this size of system, although they are expected to become more accurate for larger spin systems.\cite{Schulten78,Manolopoulos13,Lewis14,Lawrence16} In future work, we shall use the present method to analyse the results of the intriguing experiments of Wasielewski and co-workers\cite{Weiss04,Tauber06,Wasielewski06} on spin-dependent charge recombination along molecular wires,\cite{Fay17} and also show how electron spin relaxation can be incorporated in the present theory by including the molecular motions that induce it.\cite{Lindoy17}

\begin{acknowledgments}
We are grateful to Peter Hore for introducing us to spin dynamics and for many very stimulating discussions.
\end{acknowledgments}

\break

\end{document}